 \documentclass[]{emulateapj}

\usepackage{graphicx, subfigure}
\usepackage{txfonts}
\usepackage{epsfig}
\usepackage{ulem}
\usepackage{float}
\usepackage[]{natbib}
\usepackage{epstopdf} 
\usepackage{hyperref}
\usepackage{xcolor}
\newcommand{\gr}{}%cancel new command
\newcommand{\grr}{}%cancel new command
\newcommand{\grrr}{}%cancel new command

\shorttitle{HCN J=4--3 and CS J=7--6 emission from the disk around HD 142527}
\begin{document}

%% LaTeX will automatically break titles if they run longer than
%% one line. However, you may use \\ to force a line break if
%% you desire.

\title{Spatially resolved HCN J=4--3 and CS J=7--6 emission from the disk around HD 142527}

%\author{G. van der Plas\altaffilmark{1} et al.}

\author{G. van der Plas\altaffilmark{1,2}, S. Casassus\altaffilmark{1,2}, F. M\'enard\altaffilmark{2,3,4}, S. Perez\altaffilmark{1,2},  W. F. Thi\altaffilmark{3}, C. Pinte\altaffilmark{3,4}and V. Christiaens\altaffilmark{1,2}}
\affil{Departamento de Astronomia, Universidad de Chile, Casilla 36-D, Santiago, Chile}
\affil{Millenium Nucleus ÒProtoplanetary Disks in ALMA Early ScienceÓ, Universidad de Chile, Casilla 36-D, Santiago, Chile }
%\affil{UJF-Grenoble 1 / CNRS-INSU, Institut de Plan\'etologie et d'Astrophyisque (IPAG) UMR 5274, 38041 Grenoble, France }
\affil{Univ. Grenoble Alpes, IPAG, F-38000 Grenoble, France
 CNRS, IPAG, F-38000 Grenoble, France}
\affil{UMI Ð FCA, CNRS / INSU France, and Dept. de Astronomia y Obs. Astronomico Nacional, Universidad de Chile, Casilla 36-D, Correo Central, Santiago, Chile ( UMI 3386 )}

%Context, Aims, Methods, Results, Conclusions.
%What is the context? Why are these lines interesting? What is the horseshoe? Why is this object important?
\begin{abstract}
%HD 142527, dust trap, local overdensity, chemistry
The disk around HD 142527 attracts a lot of attention, amongst others because of its resolved (sub) mm dust continuum that is concentrated into a horseshoe-shape towards the north of the star. In this manuscript we present spatially resolved ALMA detections of the HCN J=4--3 and CS J=7--6 emission lines. These lines give us a view deeper into the disk compared to the (optically thicker) CO isotopes.  %We report the detection of spatially resolved HCN J=4--3 and CS J=7--6 emission with ALMA from the disk around HD 142527.  
This is the first detection of CS J=7--6 coming from a protoplanetary disk. Both emission lines are azimuthally asymmetric and are suppressed under the horseshoe-shaped continuum emission peak. % seen towards the NE of the star.%, and the CS J=7-6 line additionally shows a near / far side asymmetry. 
 A possible mechanism to explain the decrease under the horseshoe-shaped continuum is the increased opacity coming from the higher dust concentration at the continuum peak. Lower {\gr dust and/or gas} temperatures and an optically thick radio-continuum reduce line emission by freeze-out and shielding of emission from the far side of the disk.% We f
\end{abstract}


\section{Introduction}

High resolution spatially resolved observations of protoplanetary disks with a gap or central hole, i.e., transitional disks, show that these disks often are rich in azimuthal structure. These asymmetries can be divided in two groups {\grrr that either trace the disk surface or the disk mid plane. Scattered light {\grrr \citep[e.g.][]{2006ApJ...636L.153F}} and $^{12}$CO molecular line observations, e.g. spiral arms {\grrr \citep{valentin-paper}}, trace the disk surface. Radio continuum observations trace the bulk of the dust mass in the disk mid plane. Asymmetries in the latter} are  interpreted as a clustering of
larger (mm sized) dust particles (e.g. LkHa 330, SR 21 and HD 135344B
\citep{2009ApJ...704..496B, 2014arXiv1402.0832P}, HD 142527 \citep{2013Natur.493..191C} and IRS 48 \citep{2013Sci...340.1199V}) and can be explained by the presence of local pressure maxima which `trap' the dust \citep[see e.g.][and references therein for an overview]{2013ApJ...775...17L}.  In theory, these dust traps would
correspond to local minima of the gas-to-dust ratio. Such an
atypical physical environment represents an interesting laboratory for
gas chemistry.

The disk with the best studied dust trap can be found currently around
HD 142527, a nearby \citep[145 pc,][]{2004A&A...426..151A} Herbig
Ae/Be star of spectral type F6III. This star is surrounded by a large
disk with an inclination estimated to be between $\approx$ 20$^\circ$ \citep{2011A&A...528A..91V} and 28$\pm$5$^\circ$ \citep{seba-paper} and a Position Angle (PA) of 160$^\circ$. For the analysis presented in this manuscript we adopt the latter value of the disk inclination. A large opacity gap seen {\grrr in continuum emission} both in scattered light and (sub-) mm emission separates the
inner \citep[$\approx$ 10 au, ][]{2004Natur.432..479V} and outer disk 
\citep[$\approx$140 au,][]{2006ApJ...636L.153F}. The outer disk
is rich in spiral structure when imaged in scattered light
\citep{2006ApJ...636L.153F, 2012ApJ...754L..31C, 2012A&A...546A..24R, 2013A&A...556A.123C,  2013arXiv1311.7088A} and, at larger scales, CO 
rotational lines \citep{valentin-paper}. %, suggestive of dynamical perturbation by an
%orbiting body in the disk gap \citep{ 2012ApJ...754L..31C}. 
 Using ALMA at 870 $\mu$m, the outer disk
is resolved into a horseshoe-shaped dust continuum that peaks at a PA of
35$^\circ$ at radial distance of $\approx$ 1$\arcsec$
\citep{2013Natur.493..191C, 2013arXiv1309.7400F} with a surface intensity 
 contrast ratio  of $\approx$ 30 compared to the minimum of the continuum
emission at similar radius on the other side of the star. % is consistent
%with a drop in the surface density of mm-sized grains of a factor of
%$\approx$ 300\citep{2013Natur.493..191C}. 
Various J=3-2 and J=2-1 CO isotopes have been detected
coming from the outer disk but also from within the gap, and HCO+
J=4--3 emission is seen from the outer disk with a decrement coincident with the horseshoe-shaped dust continuum and in two gap-crossing filaments {\grrr \citep{2013Natur.493..191C}}. %The gas tracers

In this manuscript we present the detection of the HCN J=4--3 and
CS J=7--6 emission lines at rest frequencies of respectively 354.50547 
and  342.88286 GHz from the disk around HD 142527. 
HCN is a molecule commonly detected from disks
\citep[e.g.][]{1997A&A...317L..55D, 2004A&A...425..955T,
  2008A&A...492..469K, 2010A&A...524A..19F, 2010ApJ...720..480O,
  2011ApJ...734...98O}. The J=4--3 transition has a critical density n$_{crit}$ of 8.5 $\times$ 10$^6$ cm$^{-3}$ \citep{2004A&A...425..955T}. 
    Its dominant destruction mechanisms in disks are accretion onto dust grains, charge 
transfer reactions with H$^+$, the ion-molecule reaction with C$^{+},$~protonation reactions with H$_3^+$, HCO$^+$, and H3O$^+$ and, higher up in the disk at $\approx$ 1.5 - 2 scale heights, photodissociation by stellar
 UV  photons \citep{2011ApJS..196...25S}. 
 
% n$_{crit, HCN43}$ = 8.5 x 10$^6$, n$_{crit, CS76}$ = 2.9 x 10$^6$ \citep[c.f. Table 6 in ][]{2004A&A...425..955T}. 
%
CS emission has been detected coming from a few disks
\citep[e.g. CS J=5--4 and J=3--2, ][]{1997A&A...317L..55D,
  2010A&A...524A..19F, 2012A&A...548A..70G}. The J =
7--6 line presented in this work is the first time that this
rotational transition is detected from a disk. CS is a gas phase reservoir
for sulfur and especially its higher rotational transitions appear to
be good tracers for dense gas because of their relatively high critical densities \citep[n$_{crit}$ = 2.9 $\times$ 10$^6$  cm$^{-3}$ for J=7--6,][]{2004A&A...425..955T}. %Its high molecular weight makes it a possible tracer of turbulent motions \citep{2012A&A...548A..70G}. 
CS destruction pathways are
photodissociation in the disk atmosphere, depletion onto dust grains,
slow endothermic oxidation reaction (CS + O $\rightarrow$ CO + S),
charge transfer with ionized hydrogen atoms, and ion-molecule
reactions with primal ions, e.g. HCO$^+$ , H$_3$O$^+$ , and H$_3^+$
\citep{2011ApJS..196...25S}.\\

\begin{figure*}[ht!]
     \begin{center}
        \subfigure{
                    \label{fig:moments0CS}
            \includegraphics[width=0.33\textwidth]{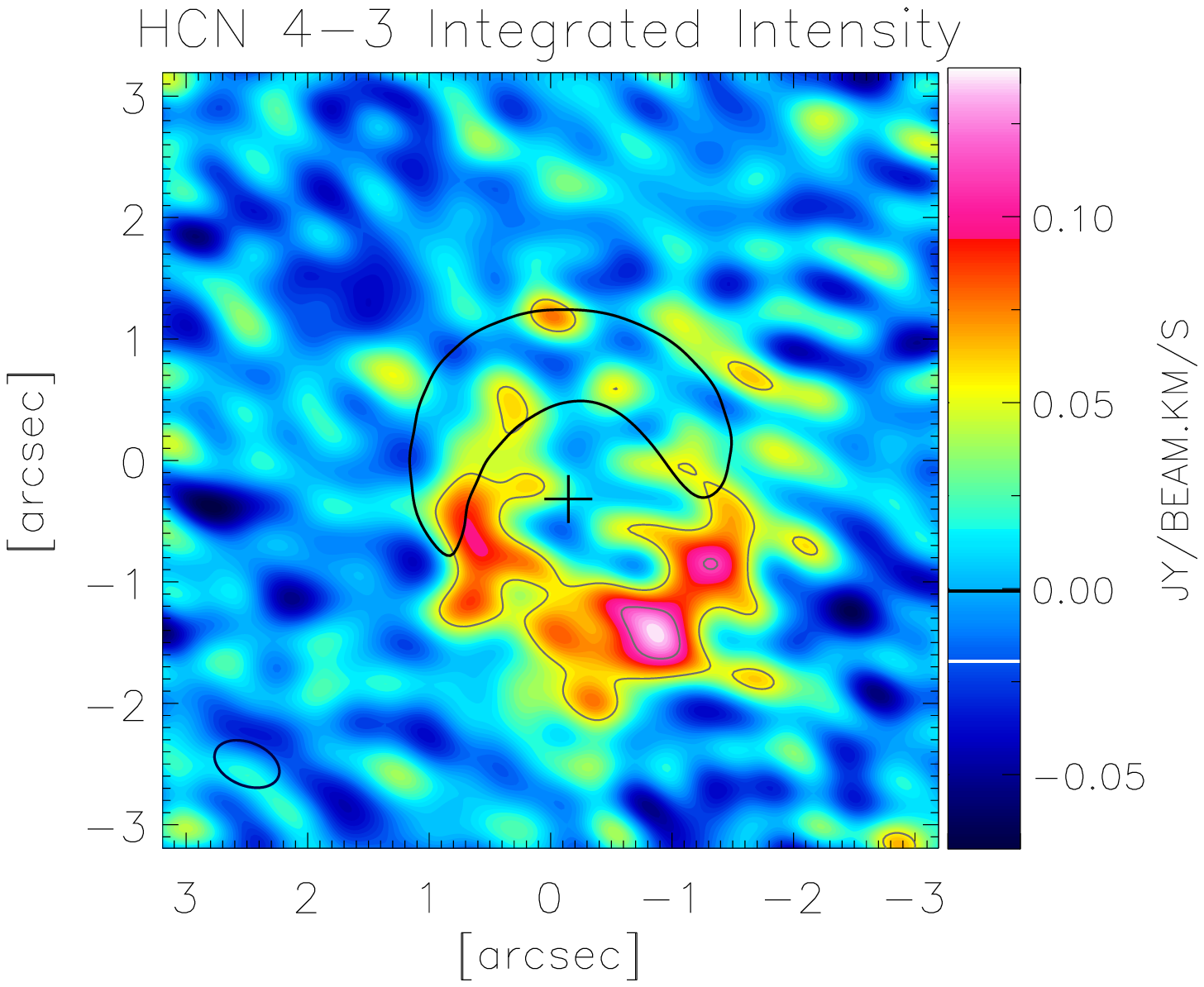}
        }%
        \subfigure{
           \label{fig:moments1CS}
           \includegraphics[width=0.33\textwidth]{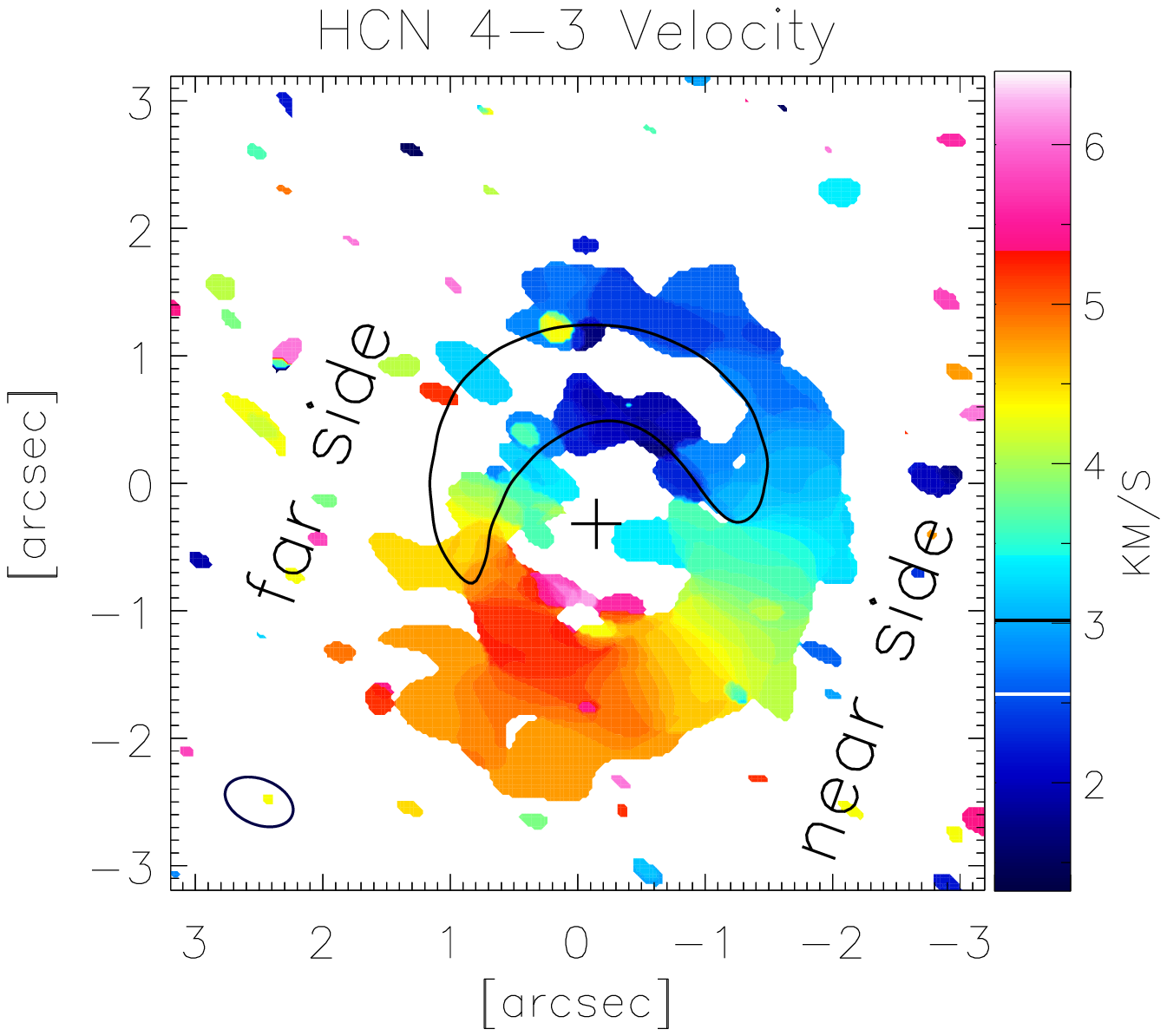}
        }%
        \subfigure{
            \label{fig:moments8CS}
            \includegraphics[width=0.33\textwidth]{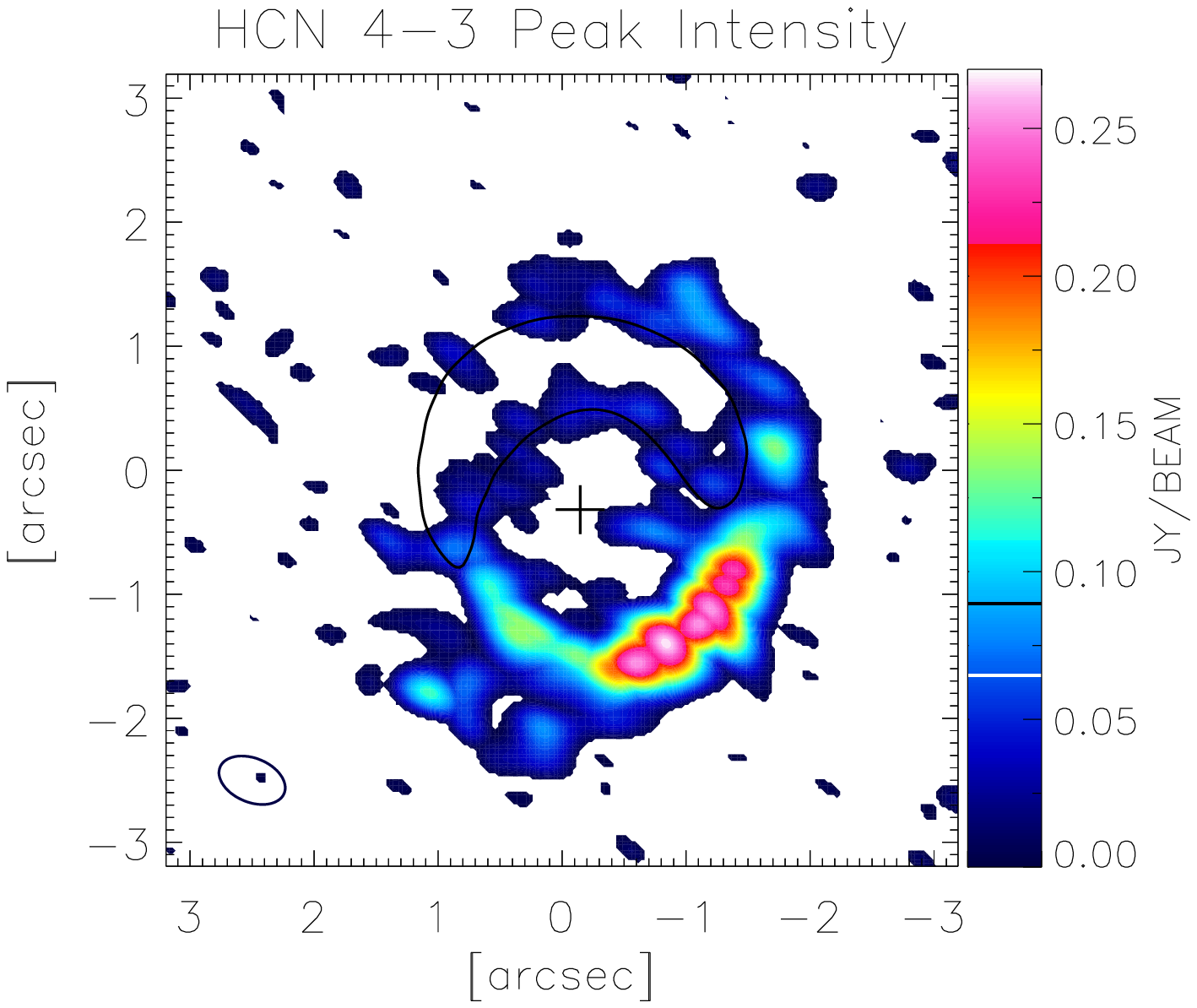}
        }%
            \end{center}
            \begin{center}
         \subfigure{
            \label{fig:moments0HCN}
            \includegraphics[width=0.33\textwidth]{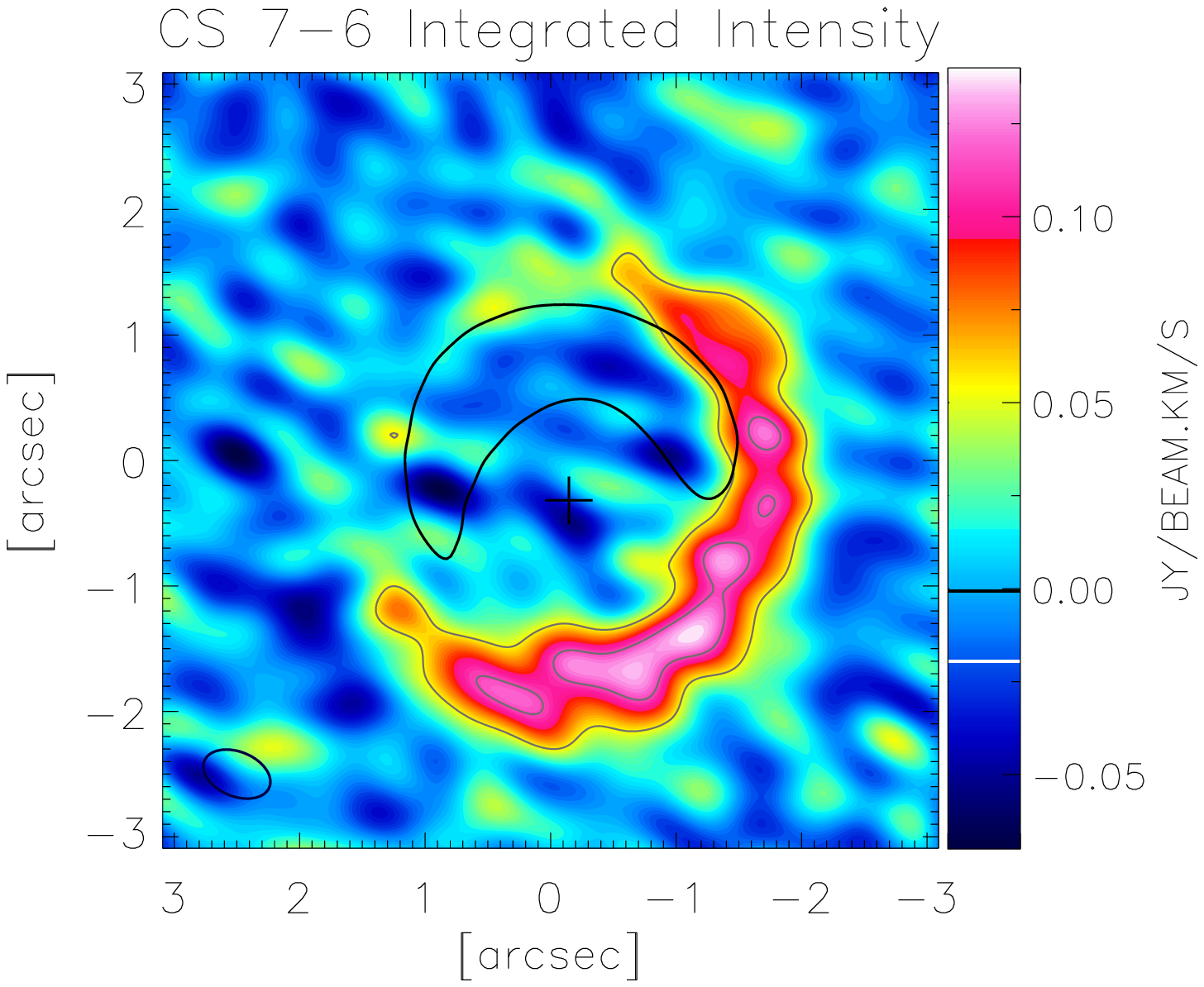}
        }%
        \subfigure{
           \label{fig:moments1HCN}
           \includegraphics[width=0.33\textwidth]{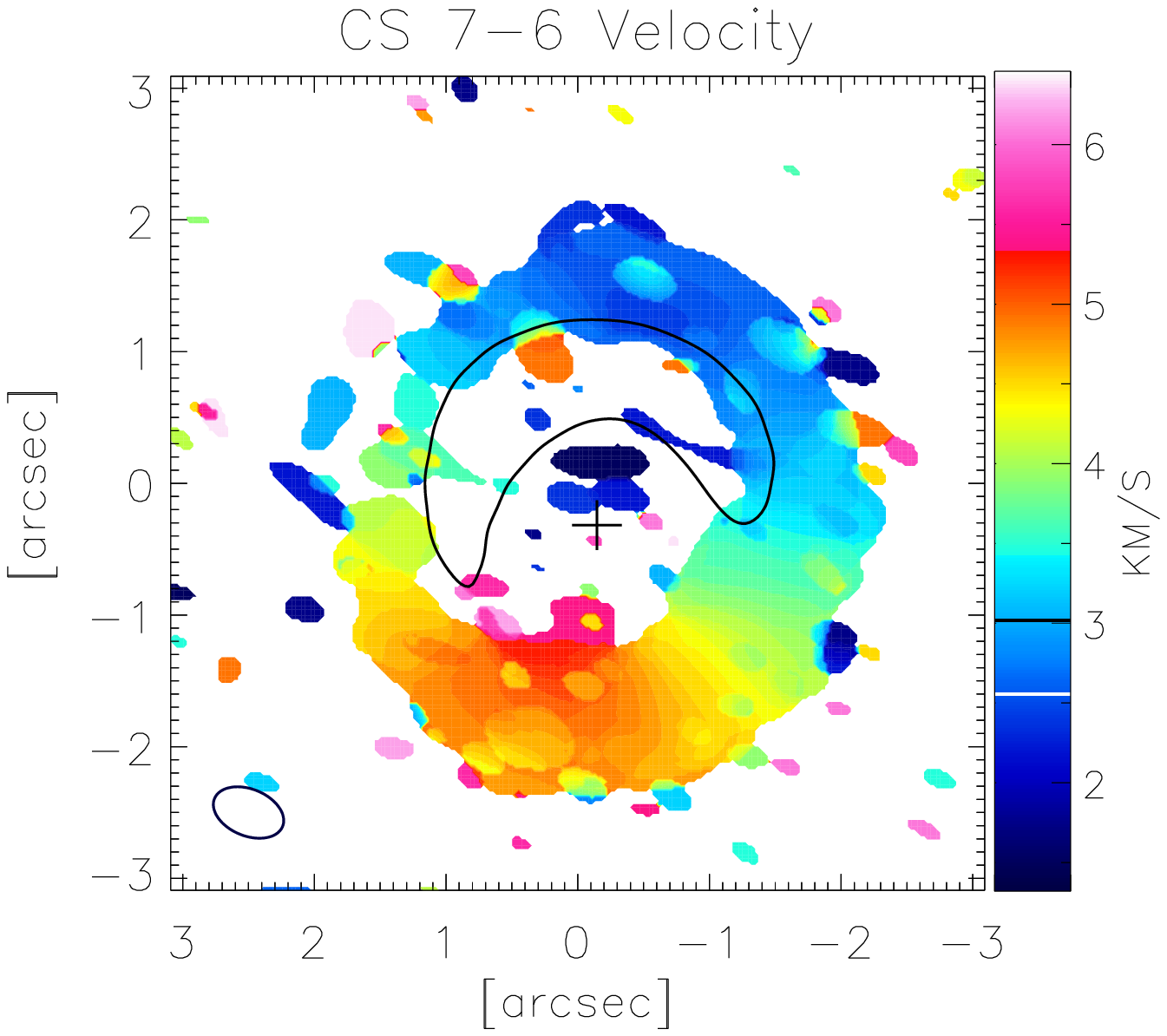}
        }%
        \subfigure{
            \label{fig:moments8HCN}
            \includegraphics[width=0.33\textwidth]{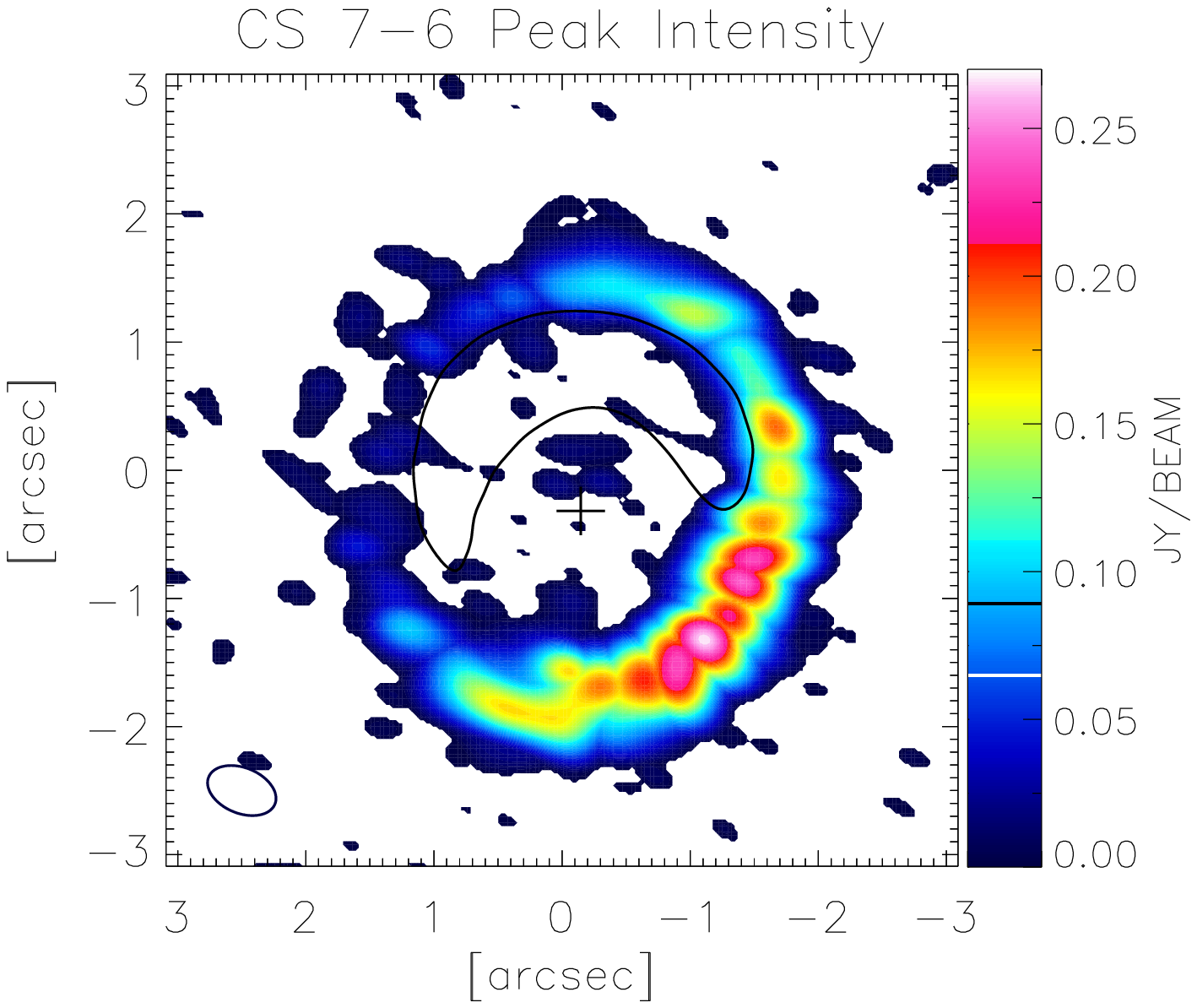}
        }%
    \end{center}
   \caption{CS J=7--6 (top row) and  HCN J=4--3 (bottom row) integrated intensity (left), velocity centroid $v_\circ$ (middle) and peak intensity $I_\mathrm{peak}$ (right). {\gr The $v_\circ$ and $I_\mathrm{peak}$ maps have been created using the maximum intensity of the spectrum at each location providing a $\geq$ 3 $\sigma$ signal}. Overplotted on the integrated intensity map are intensity contours spaced at 3 times the noise level of this map (respectively 0.014 and 0.018 Jy km s$^{-1}$ beam$^{-1}$ for CS J=7--6 and  HCN J=4--3).  The $v_\circ$ and $I_\mathrm{peak}$ maps are obtained using all signal above 3 times the RMS as determined from individual channels not containing line emission. {\gr The x and y axis are labeled in arcseconds, north is up and east is left}. The position of the star is given by a plus sign and the beam is shown in the bottom left. With a black line we show the dust continuum emission at 30\% of it's maximum. For clarity we mark the near and far side of the disk in the velocity centroid map of the HCN 4--3 emission.}
  \label{fig:momentsCS}
\end{figure*}

	\begin{figure*}[htp!]%[H]%tp]
  	\centering
  %f9.eps made with /figures/fig_chan.pro
 	\includegraphics[width=6in]{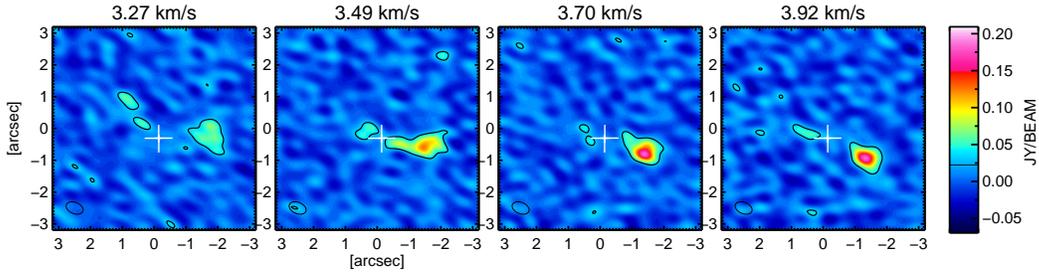}% mom 8 map used
  	\caption{Selected channel maps showing 3 channels of the HCN 4--3 emission within the gap near the systemic velocity of 3.6 km s$^{-1}$. The black contour denotes 0.2 $\times$ the peak intensity of the moment 8 map. Notation is similar to that in Figure \ref{fig:momentsCS}. }%Radial intensity profiles for all azimuths    %can be found in Figure \ref{fig:azi_slice} in the Appendix.}
  \label{fig:hcn_channels}%
\end{figure*}

\section{Observations and data reduction}\label{sec:obs+data}

For the observing strategy and data reduction we refer to \cite{2013Natur.493..191C}. 
In short summary: imaging of the CS and HCN lines was performed using the CLEAN task
 in {\tt CASA} \citep{1974A&AS...15..417H}. The data presented in this manuscript are binned in spectral 
 channels of width 0.214  km s$^{-1}$ and CLEANed to a RMS noise level of 12 mJy / 
 channel with a {\gr synthesized beam size of 0.56$\arcsec$ $\times$ 0.35$\arcsec$ and a PA of 67$^{\circ}$.}

{\gr We note that the default continuum subtraction scheme in the visibility domain left residuals which stand out as extended negatives following the horseshoe shape of the continuum in the spectral vicinity of the CS J=7--6 line.  We found no continuum subtraction artifacts in the spectral vicinity of the HCN J=4--3 line. 
  We correct for this by sampling an average continuum subtraction artifact from the spectral channels within 7 km s$^{-1}$ on either side of the affected line between respectively -6 and 1 km s$^{-1}$ and 6 and 13 km s$^{-1}$. We subtract this average from all spectral channels used in our analysis of the CS J=7--6 line.}  Removing these negatives had the effect of increasing the integrated CS J=7--6 line flux by 38\%.

  \section{Results}\label{sec:results}

\begin{figure}[htp!]
     %f7 and f8 generated using  /Users/gerritvanderplas/science/aa-package/hd142527/figures/hector/azi_four_mom8.pro
  \centering 
 \includegraphics[width=\columnwidth]{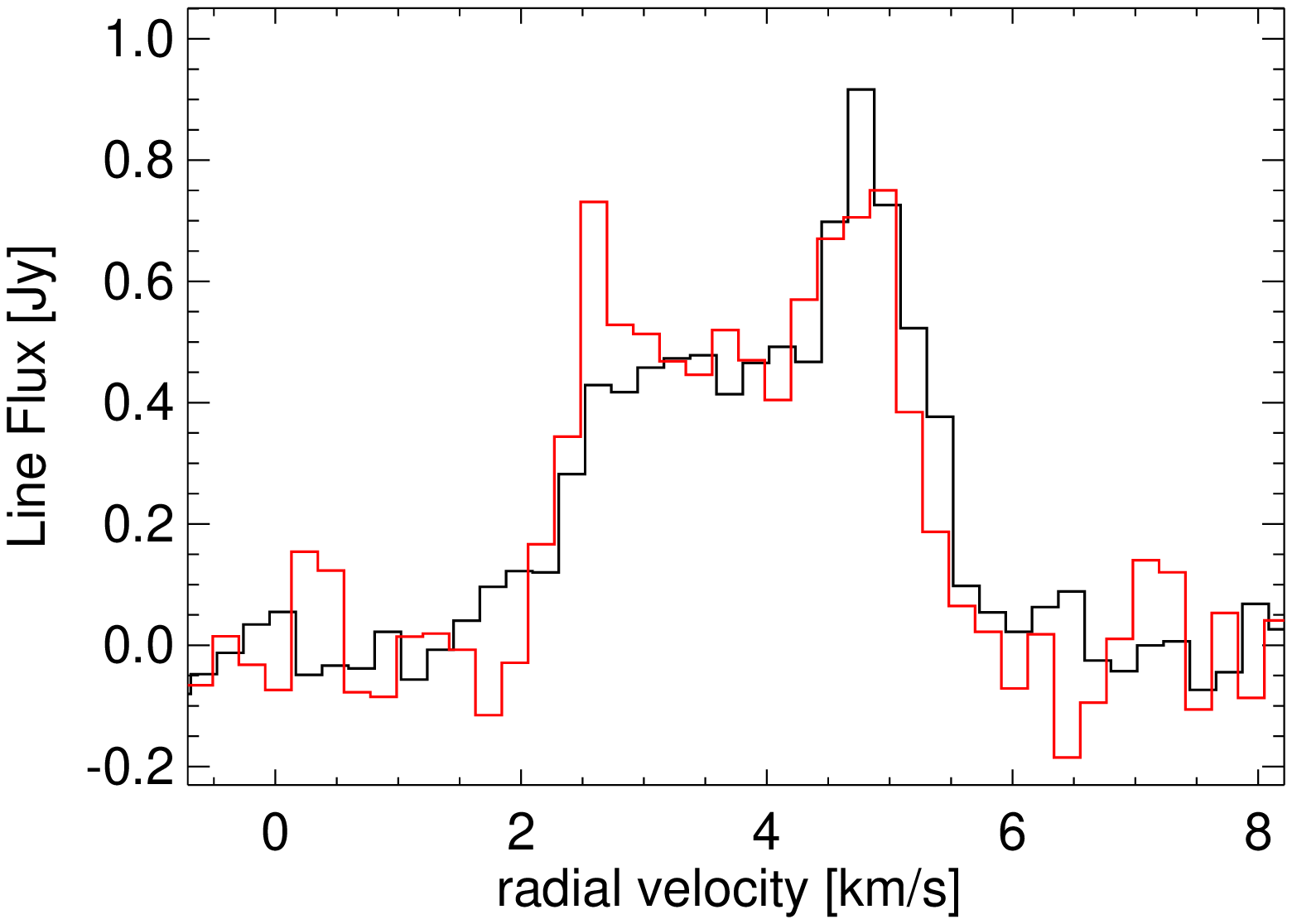} 
   \caption{HCN J=4--3 (black) and CS J=7--6 (red) emission
     lines extracted with an elliptical extraction aperture with a semi-major axis of 2.4$\arcsec$ encompassing the disk and a velocity resolution of 0.214 km s$^{-1}$ per bin. The integrated line fluxes are 1.73 $\pm$ 0.05~Jy~km~s$^{-1}$ and 1.67$\pm$0.08~Jy~km~s$^{-1}$, and the FWHM values 2.52 $\pm$ 0.21 km s$^{-1}$ and 2.71 $\pm$ 0.21 km s$^{-1}$  respectively for the HCN and CS line.}%***Perhaps include info on the extraction aperture?****}
              \label{fig:molgas}%
    \end{figure}

\begin{figure}[htp!]%[htp]
     %f7 and f8 generated using  /Users/gerritvanderplas/science/aa-package/hd142527/figures/hector/azi_four_mom8.pro
  \centering 
 \includegraphics[width=\columnwidth]{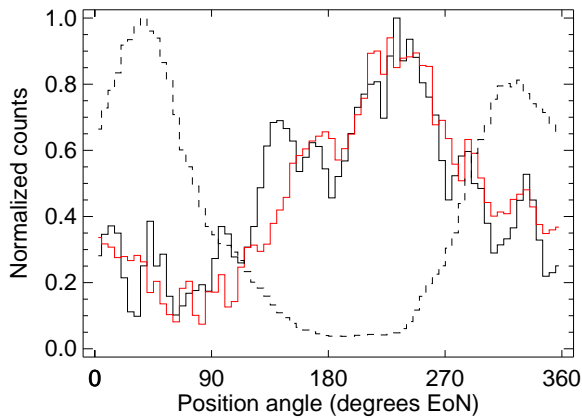} 
   \caption{Scaled HCN J=4--3 (black) and CS J=7--6 (red) azimuthal intensity profiles extracted from the peak intensity map using a ring projected to match the disk inclination and PA between 0.4$\arcsec$ and 2.2$\arcsec$ (c.f. the two right panels of Figure \ref{fig:momentsCS}). The scaled continuum intensity is shown with a dashed black line for comparison. See figure \ref{fig:azi_contrast} for the radial intensity cuts at the peak continuum and line emission.}%***Perhaps include info on the extraction aperture?****}
              \label{fig:molgas-azi}%
    \end{figure}

\begin{figure}[htp!]%%[htp]
  	\centering
  %f9.eps made with /figures/hector/fig_azicontrast_mom8_4pan.pro
% 	\includegraphics[width=\columnwidth]{./figures/f8.eps}%mom 0 map used
 	\includegraphics[width=\columnwidth]{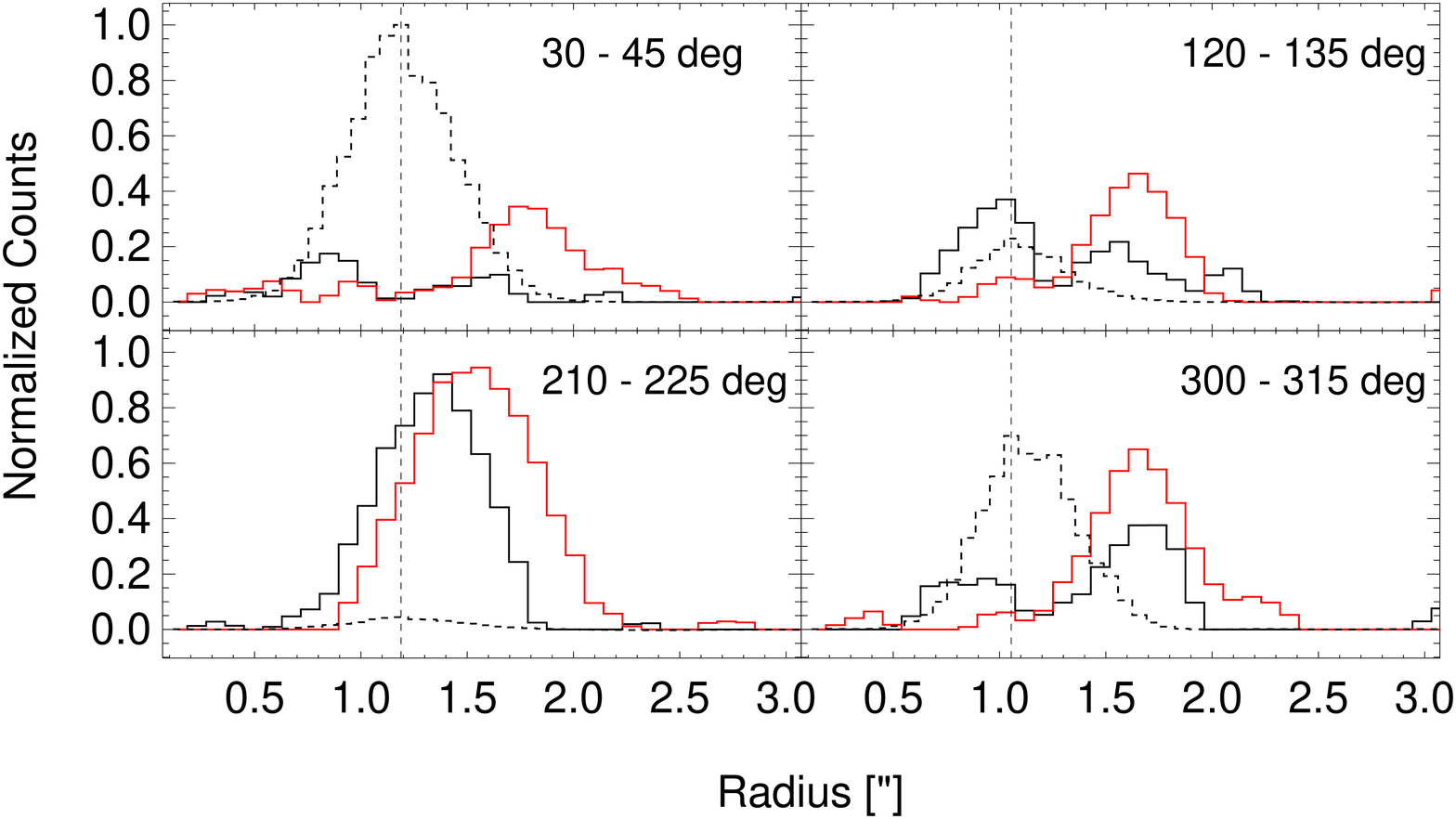}% mom 8 map used
  	\caption{Radial intensity profiles of HCN J=4--3 (black), CS J=7--6
    (red) and continuum (black, dashed) emission in four directions:
    One coincident with the peak continuum emission (top left panel, between 30$^\circ$ and 45$^\circ$ EoN), one coincident with the radially integrated peak line emission (bottom left panel, between 210$^\circ$ and 225$^\circ$ EoN), and two intermediate angles. The radial profiles have been extracted from the peak intensity maps and normalized to their   respective global maximum in the disk. We mark the radial
    location of the peak continuum emission for clarity with a
    vertical striped line.}%Radial intensity profiles for all azimuths    %can be found in Figure \ref{fig:azi_slice} in the Appendix.}
  \label{fig:azi_contrast}%
\end{figure}

We detect HCN J=4--3 and CS J=7--6 emission from the outer disk around
HD 142527. We show the CS J=7--6  and HCN J=4--3 integrated intensity,
velocity and peak intensity maps in Figure \ref{fig:momentsCS}, the integrated emission lines in Figure \ref{fig:molgas} and the azimuthal brightness variation in Figure \ref{fig:molgas-azi}. The segmented appearance of the peak intensity maps is caused by discretization of the velocity field in channel-averaged visibilities, discussed in \cite{valentin-paper}.
Both emission lines vary as function of azimuth and are suppressed in regions with strong continuum emission. The HCN J=4--3 line originates from radii smaller than the CS J=7--6 emission at all azimuths. In the outer disk, CS J=7--6 and HCN J=4--3 emission is detected beyond the horseshoe, and HCN J=4--3 marginally on its inside {\grrr as well}. The emission for both lines is strongest close to where the continuum emission is weakest (c.f. Figure \ref{fig:azi_contrast}). 

We see no counterpart for the CS and HCN emission with the spiral structure seen in scattered light and CO rotational lines. We discuss the CS and HCN emission in detail in
sections \ref{sec:results:cs} and \ref{sec:results:hcn},
respectively. 

%centering x = 0.14574, y = 0.318 arc sec from hipparcos propoer
%motion correction (seba).  We center our extraction window using the
%CO 2-1 emission \citep{seba-paper}, 0.3" towards the northeast from
%the central object %The origin of these images corresponds to the CO
%2-1 map (c.f. the plus sign in Figures \ref{fig:momentsCS} and
%\ref{fig:momentsHCN}), and is offset with xx" to the N and yy " to
%the east \citep{Seba-paper} with respect to the central star. A
%comparable offset has been reported by Fukagawa et al. 2006 based on
%... (and Simon's vizier data?). Is it significant? proper motion?,
%ALMA pointing error?  We note that the integrated

\subsection{CS J=7--6}\label{sec:results:cs}
The integrated CS J=7--6 line shows a double peaked line profile with a FWHM of 2.71 $\pm$ 0.21 km s$^{-1}$ and an integrated flux of 1.67$\pm$0.08~Jy~km~s$^{-1}$ (1~$\sigma$ ignoring calibration uncertainties). The CS emission traces the outer disk, and there is a strong asymmetry between the near and far side of the disk. %Emission from inside the gap is detected at 3 to 5 $\sigma$ confidence in 5 individual channels between 2 and 3 km s$^{-1}$. This emission is offset towards the north of the star. 
In the outer disk the emission is strongest towards the south west (PA = 200$^\circ$ - 240$^\circ$), while there is only faint emission coming from the region co-spatial to the peak continuum emission. We show the radial intensity cuts in the direction of the peak molecular and peak continuum emission in the right panel of Figure \ref{fig:azi_contrast}. 

%see file /Users/gerritvanderplas/science/DATA/ALMA/cycle0/CS76/reduced/momentmap_CS_corrected.py. at channel 142, 143, 146 (5.8, 6.0 6.6 km/s) at 3 sigma, and at channel 122,125,126 (1.5,2.2,2.4 km/s) at 5 sigma

\subsection{HCN J=4--3}\label{sec:results:hcn}

The integrated HCN J=4--3 line shape is similar to that of the CS line, but only shows the red peak. The line FWHM is 2.52 $\pm$ 0.21 km s$^{-1}$ and has an integrated flux of 1.73 $\pm$ 0.05 Jy km
s$^{-1}$ (1$\sigma$ ignoring calibration uncertainties). Compared to the CS J=7--6 line, it traces the
disk up to about the same outer radius but a smaller inner radius. The
radially integrated emission is strongest towards the south west
(between 200$^\circ$ and 240$^\circ$), and is weakest where the
continuum emission peaks. Emission inside the disk gap is detected at
4 and 6~$\sigma$ confidence levels to the east and west of the star
respectively in individual channel maps between 3 and 3.4 km s$^{-1}$, at similar velocities and location as the gap-crossing streams visible in 
the HCO+ J=4--3 reported in \cite{2013Natur.493..191C}. The radial 
intensity cuts in the direction of the peak molecular and peak 
continuum emission are shown in Figure \ref{fig:azi_contrast}. 

We note that \citet{2011ApJ...734...98O} report a 2$\sigma$ upper limit of 
 0.77~Jy~km~s$^{-1}$) on the HCN J=3--2 emission line. Assuming 
 that the HCN J=4--3 and J=3--2 emission lines are emitted from the 
 same region, optically thin and in LTE, the HCN  J=3--2 2~$\sigma$ 
 upper limit is consistent with the HCN J=4--3 detection if the excitation
  temperature is T$_{ex}~\geq$ 30 K.%, and that this upper limit is inconsistent with LTE excitation conditions and our

%	\begin{figure*}[htp!]%[H]%tp]
%  	\centering
  %f9.eps made with /figures/fig_chan.pro
% 	\includegraphics[width=6in]{f10.eps}% mom 8 map used
%  	\caption{Selected channel maps showing 3 channels of the HCN 4--3 emission within the gap near the systemic velocity of 3.6 km s$^{-1}$. The black contour denotes 0.2 $\times$ the peak intensity of the moment 8 map. Notation is similar to that in Figure \ref{fig:momentsCS}. }%Radial intensity profiles for all azimuths    %can be found in Figure \ref{fig:azi_slice} in the Appendix.}
%  \label{fig:hcn_channels}%
%\end{figure*}

\section{Discussion}\label{sec:discussion}

Both the brightness of the eastern side of the disk in thermal mid-IR emission \citep{2006ApJ...644L.133F, 2011A&A...528A..91V}  and the trailing spiral arm towards the west \citep{2006ApJ...636L.153F, 2012ApJ...754L..31C, valentin-paper} suggest that the far face of the disk is located towards the ENE. This orientation and the fact that the CS emission originates from distances larger than the inner rim (c.f. Figure \ref{fig:azi_contrast}) rule out emission from the inner rim of the outer disk inclined into the line of sight as source for the brightness asymmetry most clearly seen in the CS emission.

In the following discussion, we assume that the horseshoe-shaped continuum emission from the disk around HD~142527 is caused by dust being trapped, e.g. in a vortex, which leads to a locally enhanced dust to gas ratio {\gr and a different dust size distribution} {\grr \citep{2013A&A...550L...8B}}. The lack of CS and HCN emission under the horseshoe can be qualitatively understood when considering the implications of these different local dust properties on the presence of gas-phase CS and HCN molecules
in the dust trap. The dust provides extra surface for surface-chemistry
reactions and modifies the local temperature and radiation
field. Additionally, the dust trap may be optically thick even in the sub-mm continuum and {\grrr have a noticeable effect on the molecular emission line}. %thus  block emission below the $\tau_{850 \mu m}$ = 1 surface, provided that the molecular emission lines do not saturate above this surface.

\subsection{Temperature and chemistry}

{\gr \textbf{Within the dust trap}. A pressure maximum preferentially traps larger grains. This size sorting has the effect of increasing the local average grain size which in turn drives a lower dust temperature {\grrr because the ratio of the  mass opacity  $\kappa_{\nu}$ in the optical  and far-IR is lower for larger grains} \citep{1993Icar..106...20M}. %.of the lower mass opacity  $\kappa_{\nu}$ at shorter wavelengths

The impact of an increased dust-to-gas ratio on gas temperature is not straightforward to quantify. Dust influences the gas temperature both through cooling by thermal accommodation and through heating by photo-electric heating. Increased dust-to-gas ratios better thermalize the gas which can decrease the reservoir of thermally decoupled gas, and also plays a role in the chemistry balance. Higher dust abundances provide increased attenuation of the UV radiation field and thus shielding against photo-dissociation and photo-desorption. {\grrr An} enhanced dust abundance furthermore provides extra surface for chemical processes such as the freeze-out of CO, H$_2$O, CS and HCN molecules. Modeling the effect of a dust trap on the local gas temperature is beyond the scope of this manuscript. {\grrr For a discussion on the effect of increasing the dust to gas ratio on the gas temperature see Chapter 6 of \citet{2010PhDT........79V}, who find that increasing the dust to gas mass ratio with a factor of 10 led to a decrease in gas temperature and reduced the amount of thermally decoupled (from the dust, i.e. gas which is possibly contributing to line emission) CO gas by a factor of 3.}

%This work investigates the influence of varying the dust-to-gas ratio and PAH abundance on, amongst others, the gas temperature and the amount of thermally decoupled gas in a grid of Herbig Ae/Be disks using the thermo-chemical disk code ProDiMo \citep{2009A&A...501..383W, 2010A&A...510A..18K}. In this work, increasing the dust to gas mass ratio with a factor of 10 led to a decrease in gas temperature and reduced the amount of thermally decoupled (from the dust, i.e. gas which is possibly contributing to line emission) CO gas by a factor of 3. These results imply that there is a smaller and cooler reservoir of HCN and CS gas available to contribute to line emission and thus a higher dust to gas ratio could suppress line emission.

Observations that support this interpretations are the local minima observed in the $^{13}$CO and C$^{18}$O J = 2-1 isotopologues co-spatial to the horseshoe shaped dust continuum emission \citep{seba-paper}, which can be interpreted as a lower temperature at the line-forming region or the freeze-out of CO molecules.  Abundant crystalline water ice has also been detected from the outer disk  \citep{1999A&A...345..181M, 2009ApJ...690L.110H}. CS  and HCN in disks both freeze out at temperatures of $\approx$ 50 to 60K \citep{2006A&A...457..927G}, between the condensation temperatures of CO and H$_2$O.  {\grrr  Based on the peak of the SED at 60 $\mu$m in the ISO spectrum published by \citet{1999A&A...345..181M} we estimate the dust temperature inside the dust trap, which emits the bulk of the (sub-)mm emission,  to be $\approx$21 K.}

%We estimate the dust temperature inside the dust trap, which emits the bulk of the (sub-)mm emission,  to be $\approx$ 20 K based on the peak of the disk SED at 60 $\mu$m \citet{1999A&A...345..181M}. 
}

\textbf{Outside of the dust trap}, lower opacities {\grrr in the UV} lead to a higher sublimation and photodissociation rate and temperature {\grrr deeper in the disk}. CS molecules can form at low scale height in those regions of the disk where water freezes out and CO
does not (T between $\approx$ 20 and 100 K for a typical disk). Water
ice traps gas-phase oxygen atoms and prevents the formation of CO,
which in turn frees up carbon atoms to form reduced species such as CH and CH$_2$. These molecules can react with sulfur atoms to form
CS. HCN molecules have less stringent formation conditions and can
reach detectable concentrations higher up in the disk and closer to the
star compared to CS molecules {\gr \citep{2010ApJ...722.1607W}}.% Typically, HCN is abundant in the warm molecular disk layer above the H$_2$O ice where temperatures exceed 100K.

\subsection{Optically thick continuum}
Continuum opacity effects - a (partially) optically thick outer disk - are an 
alternative explanation of the attenuated molecular emission under the dust  trap. HD 142527 is extremely bright in the sub-mm and to fit the observed (sub-) mm emission a dust mass of 1.0 $\times$ 10$^{-3}$ M$_{\odot}$ is  required. {\grr This model has a maximum opacity at 850 $\mu$m of $\tau$ = 2.76 at the inner rim of the outer disk. \citep[private communication,][]{2011A&A...528A..91V}.}
%{\gr By assuming a surface density profile with an exponent of -1,  standard dust properties resulting in an opacity at 345 GHz of $\kappa_\mathrm{abs}$ = 0.95 cm$^2$ per gram} of dust and an azimuthally symmetric outer disk between 130 and 300 au this amount of dust translates into a ring of optically thick continuum emission at 860 $\mu$m with a maximum $\tau$ of 1.09 between 130 and 141 au. 
{\gr Given the peak flux of 0.36 Jy/beam and beam size of 0.51$\arcsec$ $\times$ 0.33$\arcsec$ \citep{2013Natur.493..191C}, the Rayleigh-Jeans equivalent brightness temperature of the peak dust emission is $\approx$20 K {\grrr and similar to our estimate of the dust temperature. The band 7 continuum is of similar size of the beam and, if radially unresolved, likely to be optically thick.}}

%To be in agreement with optically thick continuum emission the radial extent of the dust emission has to be less than 20/35 of the beam.}

To check whether shadowing by such an axisymmetric inclined optically thick ring can produce asymmetric features {\gr caused by the interception of molecular line emission from the far side of the disk in the line-of-sight towards the observer} such as e.g. the near/far asymmetry as seen in the CS emission maps, we use the radiative transfer model code MCFOST \citep{2006A&A...459..797P, 2009A&A...498..967P}. With a disk model resembling the disk architecture of HD 142527 (the same disk inclination, PA, and mass, consisting of an inner disk, a disk gap and an outer disk) we explore a range of models for which, in the inner rim of the outer disk, we vary the dust mass m$_{dust}$, the CS abundance $\epsilon_{\mathrm{CS}}$ and the dust to gas ratio $\delta$. Within a  parameter space where we vary m$_{dust}$ in the inner rim (between 5 and 30 au wide) between 6.6e-4 $\times$ M$_{\odot}$ and 4.5e-2 $\times$ M$_{\odot}$,  $\epsilon_{\mathrm{CS}}$ between 3.5e-9 and 3.5 e-11, and $\delta$ between 1 and 100, we fail to reproduce emission maps in which the near side of the inner edge of the outer disk is brighter than the far side.

The concentration of dust particles in a dust trap locally instead of in an axisymmetric ring enhances the local optical depth further, which in turn can be invoked to explain the horseshoe-shaped decrement observed in the HCN and HCO+ J=4-3 and the CO J=2-1 lines.%, but fail to explain the bulk CS7-6
The emission on the far side of the optically thick regions is absorbed,
while only excess temperature relative to the cold background leads to
emission on the near side.  Thus an optically thick horseshoe will
impart a similarly shaped decrement in line emission maps seen
face-on, with some modulation depending on the vertical temperature
structure and on the vertical location of the $\tau$ = 1 surface.

\section{Conclusions}\label{sec:conclusion}
%For CS J=7--6 this is the first detection of this rotational transition from a circumstellar disk. 
We report the detection of HCN J=4--3 and CS J=7--6 emission from the
disk around HD 142527. The emission of both molecules is azimutally asymmetric, peaks towards the southwest of  the disk and is suppressed in those regions with strong continuum emission. We discard a line-of-sight projection of the inner rim of the outer disk as origin of this asymmetry. {\grrr We suggest two possible mechanisms that can drive this asymmetry related to the presence of a dust trap: [1] The {\gr flatter dust size distribution} which drives a lower dust temperature. This in turn can lower the gas temperature and thus emissivity, and promote freeze out of CS and HCN molecules. [2] }A higher (possibly optically thick in sub-mm wavelengths) continuum optical depth associated with increased dust concentrations which quenches line emission.% and [2] a locally optically thick outer disk.

\acknowledgments
{\grr We thank the anonymous referee for the suggestions which helped make this manuscript more clear and concise, and Michiel Min for sharing his model of HD 142527.} GvdP, SC, SP, FM and VC acknowledge support from the Millennium Science Initiative (Chilean Ministry of Economy), through grant ``Nucleus P10-022-F.  GVDP, SP and SC acknowledge financial support provided by FONDECYT following grants 3140393, 3140601 and 1130949. WFT + FM acknowledge funding from the EU FP7-2011 under Grant Agreement nr. 284405. We further acknowledge funding from the European Commission's 7$^\mathrm{th}$ Framework Program
(contract PERG06-GA-2009-256513) and from the Agence Nationale pour la Recherche (ANR) of France under contract
ANR-2010-JCJC-0504-01. This paper makes use of the following ALMA data:\\ {\tt ADS/JAO.ALMA\#2011.0.00465.S}. ALMA is a partnership of ESO, NSF, NINS, NRC, NSC, and ASIAA. The Joint ALMA observatory is operated by ESO,AUI/ NRAO, and NAOJ.

{\it Facilities:} \facility{ALMA}

    \end{document}